\documentstyle{article}     % file h-matrix.tex in plain LaTeX
\begin{document}
\begin{flushright}December $18^{th}$, 1996\\
1st revision: January $5^{th}$, 1997\\
2nd revision: January $26^{th}$, 1997
\end{flushright}

{\qquad } \vspace{9mm}
 \begin{center}
 {\Large Quaternions and M(atrix) theory\\
 in spaces with boundaries
\quad\\ \quad\\ \quad\\
Lubo\v s Motl\\
\quad\\ \quad \\ 
{\it Faculty of Mathematics and Physics}\\
{\it at the Charles University in Prague,}\\
{\it Czech republic}\\
}
\end{center}
\vspace{8mm}

%\maketitle
\def \eq#1{\begin{equation}#1\end{equation}}
\def \tb#1{\left(\begin{array}#1\end{array}\right)}
\def \abs#1{\left|#1\right|}
\def \bra#1{\left\langle #1\right\vert}
\def \ket#1{\left\vert #1\right\rangle}
\def \exp{\mbox{exp}}
\def \tp{\mbox{tp\,}}
\def \ignoruj#1{}

\begin{center}
{\Large \bf Abstract of this paper hep-th/9612198}
\end{center}
%originally h-e-p---t-h-/-y-y-m-m-n-n-n

A proposal for the matrix model formulation of the M-theory on a space
with boundary is given. A general machinery for modding out a symmetry
in M(atrix) theory is used for a $Z_2$ symmetry changing the sign of
the $X_1$ coordinate. The construction causes the elements of matrices
to be equivalent to real $2\times 2$ real blocks or
quaternions and the symmetry $U(2N)$ of the original
model is reduced to $O(2N)$ or $USp(2N)=U(N,H)$.
We also show that membranes end
on the boundary of the spacetime correctly in this construction.
%------------------------------------------------------------------%

\vspace{11mm}

%\begin{center}
%{\Large This paper is dedicated do \v S\'arka.}
%\end{center}

\vspace{10mm}

\begin{tabular}{c}
\hline
E-mail: {\tt motl@menza.mff.cuni.cz}\\
WWW-page: {\tt http://www.kolej.mff.cuni.cz/}
$\tilde{\,\,}${\tt lumo/e.htm}\\
\end{tabular}
%--------------------------------------------------------------%

\newpage

\section*{Contents}

\begin{enumerate}
\item{Introduction}
\item{Modding out a symmetry in superstring theories}
\item{Modding out a symmetry in M(atrix) theory}
\item{The symmetry reversing spacetime and the membrane}
 \begin{itemize}
   \item Where do the membranes end?
   \item M\"obius and Klein bottle membranes
 \end{itemize}

\item{Gauging this symmetry and the appearance of quaternions}
  \begin{itemize}
   \item A short description of the $U(1,H)$ system
   \item The orthogonal choice
  \end{itemize}

\item{Conclusions}
\end{enumerate}

%\vspace{1cm}

\section{Introduction}

Recently the first candidate for a non-perturbative formulation
of the theory underlying all the superstring theories, branes, dualities,
D-branes and so on has been given [1] in the form of infinite-N limit
of a maximally supersymmetric matrix quantum mechanics.
This formulation offers an infinite-momentum-frame formulation of
a theory of $N$ D0-branes. Although it is not a quantum
{\it field} theory, many-particle states are contained in the Hilbert
space naturally.

This theory has $U(N)$ as its symmetry group and a modification of this
theory [2] describing type I' theory has $O(N)$ symmetry\footnote{I
write $O(N)$ and not $SO(N)$ since e.g. $(-1)$ matrix of
$O(N)$ plays a role of GSO-projection for gauge fermions [11].}
group. Apart from
unitary and orthogonal groups we know another infinite set of Cartan
compact groups: symplectic groups which can be understood as unitary
groups over the quaternions: $USp(2N)=U(N,H)$ in our notation.
As far as I know,
no matrix model of M-theory with such a symmetry has been described.

While it might be possible to investigate the M-theory from various
limits of superstring theories which contain open strings,
heterotic strings have no open strings (since left-movers and
right-movers are taken from other theories) and therefore no analogies
of D0-branes are known for heterotic strings. Therefore we could be afraid
of the possibility that just the phenomenologically most interesting
limit of the underlying theory -- namely heterotic string
or equivalently [4] M-theory on $S^1/Z_2$ -- has
not a similar non-perturbative microscopic formulation.

In this note I will try to begin to solve the drawbacks of the
last two paragraphs. Since I am just an amateur, much more work will
be necessary to correct the mistakes and to compute necessary things.

\section{Modding out a symmetry\\ in superstring theories}

Since I will use a similar technique in the case of matrix models,
let us first review the corresponding method in the superstring
perturbation theory.

The method of ``modding out'' or ``gauging''
or ``orbifolding'' a symmetry can be
used to generate various superstring models from other models.

We start with a model which has a subgroup $\Xi$ of all the
operators commuting with hamiltonian. We simply choose a group
$\Xi$ of the symmetries of the system. We will talk about $\Xi$
also as about the ``group of GSO operators''.

Now we ``identify'' elements of $\Xi$ with the identity operator.
What does it mean? It means at the first place that physical states
should be invariant under the elements of $\Xi$
\eq{\forall T\in\Xi:\qquad T\ket\psi=\ket\psi.}

But this is not the whole story. Since the shift of $\sigma$
coordinate parametrizing a string by $\pi$ is also the identity, we
must add ``twisted'' sectors where the shift of $\sigma$ is identified
with elements of $\Xi$: for each element we have one sector.
In the path-integral approach we have even more unified requirements
that we must add contributions of all the worldsheets whose
action along the noncontractible loop can be identified with
elements of $\Xi$.

The operators in $\Xi$ are defined rather formally and their
particular definition can differ sector from sector. The rules
can be supplied by some computational techniques with diverging
sums as in [5] but rules of modular invariance must be obeyed.
Modular invariance is a technicality in perturbation superstring
theory which has lost its fundamental meaning but now we are
entering to the age of a new formulation of the underlying
theory where new technicalities may become important.

Although this is quite trivial, let us note different situations
which are described by the same idea written above. We may take
a usual group of GSO operators counting numbers of some fermionic
operators. For instance, if the operator $T$ changes phases of
complex fermions $f_i$ according to
\eq{T\cdot f_i=f_i\cdot T\cdot e^{i\phi_i},}
where $\phi_i$ are angles, usually from the set $0,\pm\pi/2,\pm\pi$
(values $0,\pm\pi$ are possible even for real fermions),
the constraints of the invariance under these symmetries are the
usual GSO projections.

What happens in the twisted sectors? The $\sigma$-shift by $\pi$
is an operator we will call $\Sigma$ and its role is following one:
\eq{\Sigma\cdot L(0)\cdot \Sigma^{-1}=L(\pi),}
where $L$ is an operator being function of $\sigma$. If we identify
the operator $\Sigma$ with the $T\in\Xi$ defined above, we get
simply sectors with different boundary phases of fermions:
\eq{\Sigma\cdot f_i(0)\cdot \Sigma^{-1}=f_i(\pi)=T\cdot
 f_i(0)\cdot T^{-1}=e^{i\phi_i}f_i(0).}
There is a large industry of superstring model building
(see [6] for instance) where the
group $\Xi$ is taken to be typically $Z_2^7\times Z_4$. Most of
these models in the fermionic formulation give three generations
of quarks and leptons, often with good quantum numbers,
and have many more phenomenological virtues
and it is hard to believe that these successes are just accidental
although they were obtained in the perturbative theory.

Quite different example of gauging a symmetry, described by the same
idea, is compactification on a circle (or more generally, on tori).
In this case we take the group $\Xi$ to be isomorphic to $Z$
and containing elements
\eq{T_n=e^{2\pi inp_i r},\quad n\in Z}
shifting a coordinate $x_i$ by $nr$. The condition of the invariance
under this group
of states simply restricts the total momenta $p_i$ to be a multiple
of $1/r$. The twisted sector for the element $T_n$ contains
$n$-times winded strings:
\eq{\Sigma\cdot X_j(0)\cdot \Sigma^{-1}=X_j(\pi)=
e^{2\pi i n p_i r}X_j(0)e^{-2\pi i n p_i r}=X_j(0)+2\pi n r\delta_{ij}.}

Next good example of this construction is hidden in orbifolds. There
are for instance $Z_3$ symmetries of a torus which can be gauged out.

The symmetry operators can be combined for example with reversion
of the $\sigma$ coordinate and we get orientifolds and so on.

Although the following case is not completely standard, in some sense
also open strings can be considered as the twisted sector corresponding
to a reversion of $\sigma$. If we take $\Xi$ to be group of the
identity and the second element $T$ reversing $\sigma$, it is quite
comprehensible that the GSO projection now restricts strings to be
unoriented. For the $T$-twisted sector the following is true:
\eq{L(\pi-\sigma)=T\cdot L(\sigma)\cdot T^{-1}=\Sigma
\cdot L(\sigma)\cdot \Sigma^{-1}=L(\pi+\sigma)}
This causes the string to go from the one end to the other
and back when we let $\sigma$ increase. Periodicity becomes $2\pi$.
By the way, this doubling of the interval for $\sigma$ to $2\pi$ is
often useful.

If we take type IIB strings and make this operation, one thing must
be added: the points $\sigma=0$ and $\sigma=\pi$ are fixed
under $T$ and special things at these points can be expected.
The novelty is the 32 possible colours of the ends. It makes
$SO(32)$ type I strings from type IIB strings. While the neccessity
of 32 D9-branes is well-established today, I will only offer
a similar thing in the present construction, which could generate
the $E_8$ symmetries of the resulting heterotic string matrix model.

\section{Modding out a symmetry in M(atrix) theory}

Let us try to find a similar group $\Xi$ of operators commuting with
the hamiltonian taken\footnote{The
sign before the squared-commutator term has been changed
since I think that the commutator of two hermitian
$X^i$'s is antihermitian so its square is {\it negatively} definite.}
 from [1]:
\eq{H=R\cdot\mbox{tr}\left\lbrace\frac{\Pi_i\Pi_i}2-\frac 14
[X^i,X^j][X^i,X^j]+\theta^T\gamma_i[\theta,X^i]\right\rbrace}
We again require the physical states to be invariant under
the elements of $\Xi$:
\eq{\forall T\in\Xi:\qquad T\ket\psi=\ket\psi.}
What will be the counterpart of the twisted sectors?
I think that they will be obtained (in the string limit)
after the following procedure whose particular example
I saw in [3] in connection with compactifications to tori.

We just enlarge $N$ -- the size of the matrices -- and we will
choose a subgroup $\Xi'$ of $U(N)$, the gauge symmetry group of the
matrix model, isomorphic to $\Xi$.
Then we identify the elements of $\Xi$ with
elements of $\Xi'$. This identification is hidden in the restriction
of matrices $X^i,\theta,\Pi$ to satisfy
\eq{T'\cdot X\cdot T'^{-1}=T\cdot X\cdot T^{-1},}
where $T\in\Xi$ and $T'\in\Xi'$ are the corresponding elements of groups.
On the left hand side there is just action of an element of the group
$U(N)$ in the adjoint representation while on the right hand side
there is the physical operation.

An example is the group $\Xi$ of operators shifting a dimension
(that we want to compactify on a circle).
\eq{T_n=e^{2\pi inp_i r},\quad n\in Z.}
Then the restriction of $X$ has the result described in [3].
(I plan to describe other applications in [8].)
We can talk about that as about putting the D0-branes to all
the identified points.

Now I can also mention that the need of the
condition for states to be invariant
under the elements of $\Xi$ is now more clear from the fact
that we have identified $\Xi$ with a subgroup of $U(N)$ -- and
physical states certainly must be invariant under all the $U(N)$.

We may also have a look what happens if we try
to identify the identical element of the physical $\Xi$
with a non-identical element of $\Xi'\subset U(N)$ (the opposite
attempt cannot succeed e.g. for circular
compactification,
giving conditions like $x=x+R$), say a diagonal matrix. Then we
constrain matrices $X,\theta$ to be block diagonal and we
obtain really non-interacting copies of the universe.

I want to mention that these ideas 
applied for the circular compactification
were first realized by Banks and his collaborators [10].
Namely, Tom Banks was the first to say that for 
achieving compactification
we should look at matrices which are gauge equivalent to
translations of themselves. After some corrections due to
E. Witten this formulation (which we believe to be correct now)
was obtained, restricting directly the configuration of matrices
instead of states as in the original proposal. Afterwards
the team [1] also realized that this prescription should
be understood also as an extrapolating of the D0-brane theory
in the compactified space from the weakly coupled string theory.

I am grateful
to T. Banks for this comment and I appologize for the possible
misunderstandings which could result from my text.

\section{The symmetry reversing spacetime and the membrane}

Now we would like to apply this method to M-theory with
{\it one} boundary, where a gauge group $E_8$ should live as
Ho\v rava and Witten showed [4]. So the group $\Xi$ will be
isomorphic to $Z_2$ containing identity and the operator
$T$ reversing one of the nine transverse coordinates,
let us choose $X^1$. Physical states should be invariant under
the action of $T$.

What does $T$ make with coordinates? It must anticommute with
$X^1$ in order to change its sign
while it should commute with $X^{2}\dots X^9$ to let them intact
(below also $T=T^{-1}$).
\eq{T\cdot X^1\cdot T^{-1}=-X^1,\qquad
T\cdot X^i\cdot T^{-1}=X^i,\quad i=2\dots 9.}
Spinors should be multiplied by\footnote{We can
choose $-\gamma^1$ instead of $\gamma^1$ but we
must choose one of these possibilities. Since
$\gamma^1$ is a chirality operator for $spin(8)$
rotating $X^2\dots X^9$, we are creating a chiral
theory.}
the gamma matrix $\gamma^1$ of
the 16-dimensional real representation of $spin(9)$:
\eq{T\cdot \theta \cdot T^{-1}=\gamma^1\theta.}
Let me mention that the $spin(9)$ gamma matrices are chosen
to be real and symmetric. I will use the unified symbol ``$\pm$''
which is ``$-$'' for $X^1$, ``$+$'' for $X^2\dots X^9$
and ``$\gamma_1$'' for $\theta$'s. (Gamma matrices have
eigenvalues $\pm 1$.)

Are the terms in the matrix model hamiltonian [1] invariant
under such an operation, changing sign of $X^1$ (and also
$P^1$) and multiplying spinors by $\gamma_1$?
While the bosonic terms proportional to $\Pi_i^2$
and $[X^i,X^j]^2$ obviously are, the fermionic term
requires a careful counting of signs:
\eq{\mbox{tr}\,\theta^T\gamma_i[\theta,X^i] \, \mapsto \,
\mbox{tr}\,\theta^T\gamma_1\gamma_i[\gamma_1\theta,\pm X^i]}
For $i=1$ the three $\gamma_1$ matrices can be reduced to one
but $X^1$ changes the sign -- so the total contribution
changes the sign.

For $i>1$ due to the anticommutation
relations $\{\gamma_i,\gamma_j\}=2\delta_{ij}$ the two
$\gamma_1$'s can be transfered to each other
($(\gamma_1)^2=1$) but it changes
the sign. Since $X^i$ is invariant, also in this case the
total contribution changes the sign.

So whole the last term changes the sign under our operation.
So our operation is not complete symmetry of the hamiltonian.
We should multiply it by some next operation under which
the first two terms are even and the last term is odd.

Such an operation exists. Let me say immediately that this
operation is transposition of all the matrices -- or equivalently
(because of their hermiticity) -- their complex conjugation.
(For operators I mean that each element of the matrices
is hermite-conjugate.)

The bosonic terms are quite obviously invariant under the transposition
of matrices. The fact that transposition changes the sign of the last term
requires a careful counting of signs. Let us write the trace using
spinor indices $\alpha,\beta$, Lorentz-vector index $i$ and
$U(N)$ indices $k,l,m$:
\eq{\theta^\alpha_{kl}\gamma_i^{\alpha\beta}
(\theta^\beta_{lm}X^i_{mk}-X^i_{lm}\theta^\beta_{mk})}
If we transpose the matrices -- which corresponds to the transposition
of their indices e.g. $k,l$, we get
\eq{\theta^\alpha_{lk}\gamma_i^{\alpha\beta}
(\theta^\beta_{ml}X^i_{km}-X^i_{ml}\theta^\beta_{km})=
\theta^\alpha_{lk}\gamma_i^{\alpha\beta}
(-\theta^\beta_{km}X^i_{ml}+X^i_{km}\theta^\beta_{ml})}
the opposite sign for the result compared to starting formula.

\subsection*{Where do the membranes end?}

Let us forget for a while the $\Xi'$ being the subgroup of
$U(N)$ and study the formula (8.2) in [1] combined with our
requirement for states to be invariant under $T$ -- the
symmetry combining transposition of matrices and reversion
of $X^1$. The formula (8.2) of [1] reads (we use it for
representing matrix $X^2$ as our example):
\eq{X^2=\sum_{m,n=1-[N/2]}^{N-[N/2]}
 Z_{mn}U^mV^n \mbox{exp}(-\pi i m n/N).}
I added the phase to symmetrize the order in which the noncommuting
operators $U,V$ are written. It has the virtue of better
properties for various conjugations (see below) and its
drawback is changing the sign after $m\to m+N$ for odd $n$
and vice versa. Nevertheless, for a low energy membranes
the contributions with $\vert mn\vert <N$ are the most important and
here the phase factor differs from $1$ only a little.

Now we require the states to be invariant under $T$. $T$ has no
effect to $X^2$ so it reduces effectively to the transposition.
Let us write a particular form of the ``clock'' and the
``shift'' operators:
\eq{
U=\tb{{cccc}1&&&\\ &e^{2\pi i/N}&&\\ &&e^{4\pi i/N}&\\ &&& \ddots},\qquad
V=\tb{{cccc}0&&\dots&1\\ 1&0&&\\ &1&0&\\ &&1&\dots}
}
Clearly, $U$ is symmetric and the transposition of $V$ is $V^{-1}$.
That means that the transposition inverts one of the matrices ($V$).
Alternatively, if we use the complex conjugation, $V$ is real
and the complex conjugate of $U$ is $U^{-1}$.

In both cases, the operation inverts one of the two matrices.
To be concrete, let us talk about the transposition. Using the
facts just stated it is easy to show that
\eq{(U^m V^n e^{-(\pi i m n / N)})^T=
e^{-(\pi i m n / N)} V^{-n}U^m=
U^mV^{-n}e^{\pi i m n/N}}
after the transposition in the contribution to the $X^2$
proportional to $Z_{mn}$ the remaining factor will be replaced
by the factor which was associated to $Z_{m,-n}$ before
the transposition. Such a changing of Fourier mode $n$
to $-n$ is in
the continuous basis equivalent to reversion of one coordinate
on the fuzzy torus-like membrane. Thus the condition for
invariance of the states i.e. for the symmetry of $X^2$ under
the transposition
(let us suppose an eigenstate of matrix elements of $X^2$ and
understand $X^2$ as a classical matrix)
tells us something like
\eq{X^2(p,-q)=X^2(p,q).}
For the $X^1$ coordinate changing the sign
included in $T$ will change the formula to
\eq{X^1(p,-q)=-X^1(p,q),}
which means that the membrane ends with its boundary
$q=0$ on $X^1=0$:\quad
$X^1(p,0)=0$. The same is true for $q=\Delta/2$ where
$\Delta$ is the period of $q$ in the fuzzy torus. Here
there is the second boundary and the torus is restricted
to a cyllinder.

Now we could be afraid of the fact that the restricting operators
$X,\theta$ will change this result. I do not think so because the
role of these restrictions of operators can be understood as
the freedom to produce the twisted sectors and we can always think
about a ``quite an isolated world'' located in the part of the
matrices where elements of $\Xi'$ look the same (in one of the
blocks).

In fact, the
argument of this section is more reliable in the orthogonal case
(using $\sigma^3$ into $T'$ below) than in the symplectic one.
Nevertheless, we have showed that using real symmetric matrices
(instead of complex hermitean ones) restricts the fuzzy torus to be
the fuzzy cyllinder. Thus I disagree with various recent claims
expressing the absence of open membranes in the
non-commuting torus construction and the need to add
some boundary terms.

\subsection*{M\"obius and Klein bottle membranes}

I studied the question a little. The transposition or the complex
conjugation produces the cyllinder, mapping
\eq{U,V \mapsto U,V^{-1}\quad \mbox{or} \quad U^{-1},V.}
We could obtain also the M\"obius strip in a similar way.
The only thing we must practice is the corresponding mapping
\eq{U,V \mapsto V,U.}
The fact that it produces a M\"obius strip is clear from the
picture I cannot put here. But the usual representation
of the strip --
the square with
a pair of distinct opposite boundaries
and the second pair antiidentified -- is obtained
here as a square of $1/2$ area compared to the torus and is
by 45 degrees rotated. If you draw this smaller square into
the original (torus) square which is divided by the
symmetry around the axis $x=y$ and move it by $\Delta/2$ above,
you will understand why it
has the M\"obius topology.

We can obtain such a mapping by a minor modification of
the conjugation $W\mapsto \bar W$ for $W=U,V$, namely
by adding a discrete Fourier transformation ($\omega=\exp(2\pi i/N)$):
\eq{W\mapsto F\cdot \bar W\cdot F^{-1}, \mbox{\quad where\quad}
F=\frac1{\sqrt N}\tb{{ccccc}
1&1&1&1&\dots\\
1&\omega&\omega^2&\omega^3&\dots\\
1&\omega^2&\omega^4&\omega^6&\dots\\
1&\omega^3&\omega^6&\omega^9&\dots\\
\vdots&\vdots&\vdots&\vdots&\ddots},}
But I do not know where this matrix could appear.
(Note that $F^2$ is an antidiagonal matrix and
$F^4=1$.)

Finally, if we would restrict the matrix representing the membrane
to be invariant under the mapping of the type
\eq{U,V\mapsto U^{-1},-V,}
we could get a membrane with the Klein bottle topology.
It differs from the cyllinder only by the minus sign in $-V$.
This change denotes a shift by $\Delta/2$ so the membrane
should be invariant under the combined operation of
reversing one coordinate and shifting the second by $\Delta/2$.
Clearly, one pair of opposite sides is still identified
and the second pair is anti-identified, giving a standard
representation of the Klein bottle.

\section{Gauging this symmetry and the appearance of quaternions}

In the previous section we were discussing a pleasant result of the
required invariance of states. Now we would like to study the
restriction of operators. We must choose an element $T'$ of $U(2N)$
(since now we take the size of matrices to be even) which will
represent $T\in\Xi$. Since also a transposition plays the game,
all the choices will not be completely equivalent. Let us
suppose that $T'$ is a tensor product of unit matrix and some
$2\times 2$ matrix whose square is also the unit matrix.
I did not want to use a trivial
one (again a unit matrix) -- so the most
natural candidates are Pauli matrices. $\sigma^1$ and $\sigma^3$
give some results
(see the ``orthogonal case'')
but the most interesting seemed to me to use the
imaginary Pauli matrix -- $\sigma^2$.
I could not say why this choice was better than others,
but I felt it from the resulting symmetry.
(Now I think that choosing $\sigma^1$ or equivalently
$\sigma^3$ gives the more interesting
and realistic
theory which I will shortly
discuss later.)
Let $T'$ be
the block diagonal
matrix consisting of $\sigma^2$'s on the block diagonal.
\eq{T'=\mbox{diag}(\sigma^2,\sigma^2,\sigma^2,\dots)}
Let $Y$ denote $X^{2\dots 9}$ or $X^1$ or $\theta$'s and
$\pm$ is minus for $X^1$, plus for the remaining $X$'s
and $\gamma_1$ for $\theta$'s. We require as in [1] all
the $Y$'s being hermitean complex matrices. Let us write
the requirement for $Y$'s:
\eq{T'\cdot Y\cdot T'^{-1}\quad =\quad \pm \bar Y,}
where $\bar Y$ means complex conjugation (i.e. hermitean conjugation
of matrix's elements). Those $\sigma^2$'s in $T'$ act
on each $2\times 2$ block of $Y$ giving a restriction for it:
\eq{\tb{{rr}0&-i\\ i&0}
\tb{{cc}A&B\\ C&D}
\tb{{rr}0&-i\\ i&0}=\tb{{rr}D&-C\\
-B&A}\quad = \quad \pm\tb{{cc}\bar A&\bar B\\
\bar C&\bar D}}
so the conditions are $D=\pm \bar A$, $-C=\pm\bar B$.
Let us suppose for a while that $\pm=+$. Then the conditions
constrain the $2\times 2$ blocks of $Y$'s to be of form
($a,b,c,d$ are real)
\eq{\tb{{rr}a+bi&c+di\\ -c+di&a-bi}}
Similarly, for $\pm=-$ we require the $2\times 2$ blocks to be
$i$ times the matrix of the type above. For spinors
$\pm=\gamma_1$ but this is nothing new: half of components
have $\pm=\gamma_1=+1$ and half of them have $-1$.
The $2\times 2$ matrices of the form above have exactly
the same multiplication rules as quaternions
$a+bi+cj+dk$. And also the hermitean conjugation of such
a matrix gives the conjugate quaternion $a-bi-cj-dk$.
Thus we can replace these blocks by quaternions.

The situation $\pm=-$ differs in one basic aspect only.
We can again consider this as a representation of
quaternions (where $i$ or $-i$ -- I
cannot decide now -- must be added to each product)
but their hermitean conjugation gives {\it minus}
conjugate quaternion $-a+bi+cj+dk$ in the same representation.

This mean that $X^2\dots X^9$ (and half spinors $\gamma_1=1$)
can be regarded as
hermitean quaternionic matrices (hermitean conjugation for
quaternionic matrices is a combination of transposition
and quaternionic conjugation) while $X^1$ (and the rest of
spinor components) as antihermitean quaternionic matrices.

In a process of a typical orbifolding, the $U(2N)$ symmetry
would be reduced to $U(N)\times U(N)$ symmetry. Now, because
of the complex conjugation, the symmetry would be restricted
to $U(N)$ in a generic case. Therefore this generic case
should be inconsistent since (intuitively) the consistent
theories should have about the same dimension of the
gauge group as one of $X_i$'s, say $X_1$.
(Note for instance that all the groups
$U(N)\times U(N)$, $O(2N)$ and $USp(2N)$ have
dimension about $2N^2$.)
Thus only extremal
choices have a chance to be consistent. Giving $\sigma^2$
to $T'$ leads to the quaternionic theory while using
$\sigma^1$ or equivalently $\sigma^3$ should
create a theory with orthogonal symmetry.
In the quaternionic case, the $T'YT'^{-1}=\bar Y$
corresponds to $jYj^{-1}$ and the invariance under
this $Z_2$ combines with the generic
$U(N)$ to the requirement of the invariance under
whole $U(N,H)$.

Now we could be surprised by the antihermitean form of $X^1$.
The remaining coordinates are correct and have real numbers
on the diagonal. But $X^1$ has ``purely quaternionic'' numbers
$bi+cj+dk$
on its diagonal. It seems as the $X^1$ coordinate exists three
times.

But fortunately, this is not the case. The reason is that
directions in the three-dimensional space of numbers $bi+cj+dk$
are all equivalent since they can be transformed to each other
(due to the noncommutativity of quaternions) by the transformations
of $USp(2N)=U(N,H)$ ($H$ denotes the set of quaternions and
the matrices $Y$'s are taken to be $N\times N$ quaternionic).
In fact, these directions are equivalent even to their opposite.
But even this should not be too big surprise since the sign
of $X^1$ coordinate is unphysical.

Even in the potential case where we would use more than one
($X^1$) antihermitean quaternionic matrix, no problem would
arise because these coordinates would have on the corresponding
sites of diagonal typically ``pure imaginary quaternions''
from the same direction - i.e. real-number-proportional
to each other. In the opposite case the energy containing
commutators would increase rapidly again due to the
noncommutativity of quaternions:
\eq{[i,j]=2k,\qquad [i,j]^2=-4}

So we just say that a formulation of the M-theory on a space
with boundary requires quaternionic matrices $X,\Pi,\theta$
where $X^1$ and half of $\theta$'s are antihermitean while
the other are hermitean and the hamiltonian looks essentially
as in [1]. Perhaps, new degrees of freedom -- perhaps
again in the fundamental representation
(natural from the point of view that it is associated to
$N$ elements of boundary of the membrane which has
$N^2$ elements because it is associated with matrix)
of the $U(N,H)$
symmetry group -- should be added as in [2]
together with terms in the hamiltonian
\eq{\sum_{r=1}^{8 or 16?}\lambda_r^{k\dagger} X^1_{kl} \lambda_r^l}
Their existence could
be explained by similar arguments concerning the transposition
of matrices as the origin of 32 colours at the ends of
type I string was explained as a side-effect of reversion of
$\sigma$.
Also, a prematrix theory might be found where elements of
$X$\dots operators would be represented as states, restrictions
of these operators as restrictions of these states and
the new $16$ $O(N)$ vectors would arise
from the fixed points of a $Z_2$ operation
in a similar way as in description of (0,1) heterotic strings by
(2,1) strings [7].
(In fact, I was trying to obtain the fields
necessary for the gauge symmetry from the original spinors
so that they would loose their Lorentz quantum numbers
but this is perhaps a lousy idea.) It is quite interesting
because some papers indicate that it should be possible
to get all the compactifications from the original M-theory
without adding degrees of freedom. Maybe that this result
is limited to theories which originate from
{\it untwisted} algebra [7] of $(2,1)$ heterotic strings.

\subsection*{A short description of the $U(1,H)$ system}

In [1], the most simple case $N=1$
with the symmetry $U(1)$ gave a free theory with
256 states having a momentum. In the quaternionic case, the
simplest case has symmetry $USp(2)=U(1,H)$ which is isomorphic
to $spin(3)$. $X^2\dots X^9$ and $\Pi^2\dots\Pi^9$ as well
as $\theta$'s with $\gamma_1=1$ are hermitean $1\times 1$
quaternionic operator-matrices - so they are hermitean scalars.
But the remaining components of $\theta$ and $X^1$ as well
as $\Pi_1$ are {\it antihermitean}, so they have the form
$bi+cj+dk$ and transform as vectors under the $spin(3)$ group.

Now we would like to see if the physics in the bulk of [1]
is reproduced in this model. Let us begin with an eigenstate
of $Y^1$ components.
We can make a $spin(3)$ transformation to achieve
\eq{Y^1_x=Y^1_y=0.}
The real physical states invariant under the $spin(3)$
are then obtained by the integration over all the group.
The hamiltonian looks like
\eq{H=\sum_{i=2}^9 \frac{\Pi_i\cdot\Pi_i}2
+\sum_{j=x,y,z}\frac{\Pi_1^j\cdot\Pi_1^j}2
+\sum_{r=1}^8 i\varepsilon_{ijk}\cdot X^1_i
\theta^r_j\theta^r_k.}
The momenta $2\dots 9$ contribute to the energy in the same
way as in [1] and the same is true also for $\Pi_1$ when
$\abs{Y^1}$ is large. The components of $\theta$ with
$\gamma_1=+1$ are not contained in hamiltonian --
in the same fashion as in [1] where they ensure
(together with the $\gamma_1=-1$ components)
the 256 degeneracy of states.

Where are the $\gamma_1=-1$ components in our construction?
Because of the $\varepsilon_{ijk}$, for $Y^1$ having
the $z$-direction
the $z$-components of spinors with $\gamma^1=-1$ decouple
from the hamiltonian and just these states play the role
of the scalar $\theta$ components with $\gamma^1=-1$ in [1].

But the $x$ and $y$ components of $\theta_{\gamma_1=-1}$
are interacting. It is natural to combine them to combinations
$\theta_x\pm i\theta_y$. The ground level is anihilated by
all $\theta_x+i\theta_y$ (or minus?) and is a Lorentz scalar
since the sum of weights in any representation equals zero.
But such a ground level is not a $spin(3)$ scalar since
it has $j_z=-4$. (The opposite ground level obtained
by application of all the eight creation operators
$\theta_x-i\theta_y$ must have $j^z$ greater by 8 and
at the same moment, inverse to the $j_z$ of the true
ground level.)

So there is an anomaly. Its result is by the way also an
energy proportional to $Y^1$. I think that the most natural
way to cancel this anomaly is to add fermions in the fundamental
representation of $spin(3)$ i.e. spinors together
with a hamiltonian term like
\eq{\sum_{s=1}^8\sum_{p,p'=1,2}
\lambda^{\dagger p}_s\sigma_{pp'}^i X^1_i\lambda_s^{p'}}
We need $8$ such
spinors because each of these have two components but they
have only $j_z=\pm 1/2$. I
have thought for a time that these fermions
can be the source of the $E_8$ symmetry but I found that
they can generate symplectic symmetries much more easily
than $SO(16)$\dots

The spin-statistics theorem is obeyed because of the
$j=0$ condition: the only variable which could break it
($\lambda$) has $j=1/2$ so its creation operators must
be always paired. Nevertheless, I do not know if all these
ideas can lead to a really consistent theory\dots Only
now I realized that maybe it's more natural to add bosons
(and not fermions $\lambda$) but I leave this question to
a future work.

\subsection*{The orthogonal choice}

I just repeat the discussion
from the last section for $\sigma^1$:

Those $\sigma^1$'s in $T'$ act
on each $2\times 2$ block of $Y$ giving a restriction for it:
\eq{\tb{{rr}0&1\\ 1&0}
\tb{{cc}A&B\\ C&D}
\tb{{rr}0&1\\ 1&0}=\tb{{rr}D&C\\
B&A}\quad = \quad \pm\tb{{cc}\bar A&\bar B\\
\bar C&\bar D}}
so the conditions are $D=\pm \bar A$, $C=\pm\bar B$.
Let us suppose for a while that $\pm=+$. Then the conditions
constrain the $2\times 2$ blocks of $Y$'s to be of form
($a,b,c,d$ are real)
\eq{\tb{{rr}a+bi&c+di\\ c-di&a-bi}}
Similarly, for $\pm=-$ we require the $2\times 2$ blocks to be
$i$ times the matrix of the type above. For spinors
$\pm=\gamma_1$ but this is nothing new: half of components
have $\pm=\gamma_1=+1$ and half of them have $-1$.

The matrix above can be written as ($a,b,c,d$ are real)
$a+ib\sigma^3+c\sigma^1-d\sigma^2$
and is equivalent to
$a+ib\sigma^2+c\sigma^x+d\sigma^3$
which is real. Note also that the hermitean conjugation
makes the same operation in both cases: $b\mapsto -b$.

A similar result we would get for $\sigma^3$ and in fact
also $1_{(2\times 2)}$ gives real blocks.
In the latter case we see clearly that $X^1,\Pi^1$
and $\gamma_1=-1$ spinors are antisymmetric
purely imaginary\footnote{Multiplying by $i$ gives
real antisymmetric matrices. We keep the convention
of hermiticity for all the matrices.}
matrices while the others are symmetric
real as in [2].
So these
prescriptions require $X$'s and $\theta$'s to be real
(elements to be hermitean) and the symmetry $U(2N)$
is restricted to $O(2N)$. Now I think that this orthogonal
case may lead to a standard $E_8$ symmetry on the boundary
after adding a vector of $SO(16)$ tensored with
a vector of $O(2N)$. In fact I see no differences between
the recent model and the model of [2].
So I can say the most visible result of [2] that the states
$120$ are represented in the $O(2N)$ system while the
remaining states
$128$ of $E_8$ are included in the $O(2N+1)$
(which is possible if we choose unit matrix instead
of Pauli ones). For instance,
for the $O(1)$ system, all the antisymmetric matrices
($X^1,\Pi^1$ and spinor components with $\gamma_1=-1$)
must equal to zero -- so the states are living at $X^1=0$ --
and only operators $X^i,\Pi_i$, $i=2\dots 9$ and spinor components
of $\gamma_1=+1$ plays the role, giving the standard
$16_{from\,spinors}\times 256_{from\,\lambda's}/2_{from
\,the\,projection}$ degeneracy of the states.
I plan a paper [11] on bosonic $E_8$ and some
corrections of [2].

\section{Conclusions}

In this note I was trying to obtain a matrix model formulation for
M-theory on a space with boundary.
The orbifolding of $Z_2$ symmetry seems to give two
apparently consistent possibilities.
The orthogonal one restricts matrices to be equivalent
to real matrices and gives $O(2N)$ symmetry
(this choice was forgotten in the first version of the paper)
or even $O(N)$ symmetry not only for even $N$'s.
Now I believe that just
the missed orthogonal case describes the
M-theory with $E_8$ on one boundary [2] originally found
from type I' D0-branes in the same sense as
the full M-theory on $M^{11}$ was obtained from
type IIA D0-branes [1].

The second -- symplectic version
may give another consistent theory or even (if the added
degrees of freedom are bosons) the same theory with the
bosonic representation of $E_8$.
In this one quaternionic matrices appeared
quite naturally. This theory may give also
a different consistent
system in 11 dimensions -- with the boundary lived e.g.
by $USp(16)\times SO(8)^4$ multiplet. (Symplectic
symmetries arise in the quaternionic case much better
than the orthogonal ones.) This theory would not
be described by any limit of any string theory known.
The real and quaternionic cases have something common:
complex representations represent
a group as a subgroup of $U(N)$ and are not equivalent to their
complex conjugates while the real (subgroup of $O(N)$) and
pseudoreal=quaternionic (subgroup of $USp(2N)$) are.

It should be verified if the new physical system satisfies a
correct supersymmetry algebra [2,11,future]. Also
the idea should be generalized
to compactification to $S^1/Z^2$ [8,future];
now we were only briefly
discussing orbifolding to $R/Z^2$ where only one
e.g. $E_8$ appears.
Physical states in the ``bulk'' of the spacetime should be
the same as in [1] and new states
(super-Yang-Mills $E_8$ multiplet)
associated to the boundary should be found.

If the quaternionic theory would appear consistent, we
would have theories with all the possible gauge groups
from the infinite sets of simple groups.
Even $U(m,n)$ symmetry group has been used to study brane-antibranes
interaction and the analytical continuation from $U(m+n)$
was showed to correspond to crossing symmetry [9].

I appologize for my poor English and I wish you a M(erry) Christmas
and a Happy New Year.

%------------------------------------------------------------------
%\newpage
\section*{References}
\begin{enumerate}
 \item T.Banks, W.Fischler, S.H.Shenker, L.Susskind:
 {\it M Theory As A Matrix Model: A Conjecture,} hep-th/9610043
 \item S.Kachru, E.Silverstein:
 {\it On Gauge Bosons in the Matrix Model Approach to M Theory,}
 hep-th/9612162, the model was first written down in\newline
 U.Danielsson, G.Ferretti: {\it The Heterotic Life of the D-particle,}
\newline
 hep-th/9610082
 \item W.Taylor:
 {\it D-brane field theory on compact spaces,} hep-th/9611042
 \item P.Ho\v rava, E.Witten:
 {\it Heterotic and Type I String Dynamics from Eleven Dimensions,}
hep-th/9510209
 \item L.Motl:
 {\it Two-parametric zeta function regularization in superstring theory,}
hep-th/9510105
 \item A.Faraggi: {\it Realistic Superstring Models,} hep-ph/9405357
 \item D.Kutasov, E.Martinec: {\it New Principles
for String/Membrane Unification,} hep-th/9602049
 \item L.Motl: {\it Proposals on nonperturbative superstring
interactions,} \newline hep-th/9701025   %was in preparation
 \item V.Periwal: {\it Antibranes and crossing symmetry,}
 hep-th/9612215
 \item T.Banks: speech at Aspen Workshop, summer 1996
\newline and
 Princeton, September 1996, unpublished
 \item L.Motl: {\it Bosonic representation of gauge
 symmetry in M(atrix) theory,} hep-th eprint in preparation
\end{enumerate}
\end{document}